# The Research and Optimization of Parallel Finite Element Algorithm based on MiniFE

Meng Wu，Can Yang，Taoran Xiang，Daning Cheng

Institute of Computing Technology，UCAS，Beijing，China

**Abstract:** Finite element method (FEM) is one of the most important numerical methods in modern engineering design and analysis. Since traditional serial FEM is difficult to solve large FE problems efficiently and accurately, high-performance parallel FEM has become one of the essential way to solve practical engineering problems. Based on MiniFE program, which is released by National Energy Research Scientific Computing Center(NERSC), this work analyzes concrete steps, key computing pattern and parallel mechanism of parallel FEM. According to experimental results, this work analyzes the proportion of calculation amount of each module and concludes the main performance bottleneck of the program. Based on that, we optimize the MiniFE program on a server platform. The optimization focuses on the bottleneck of the program - SpMV kernel, and uses an efficient storage format named BCRS. Moreover, an improving plan of hybrid MPI+OpenMP programming is provided. Experimental results show that the optimized program performs better in both SpMV kernel and synchronization. It can increase the performance of the program, on average, by 8.31%.

## 0 Introduction

Finite element method(FEM) is an important numerical method in modern engineering design and analysis, and has been widely used in various industry fields such as transportation, water conservancy, construction, aerospace,etc[1]. With rapid development of modern technology, engineering structures get larger and more complicating. Thus traditional serial FEM is not able to deal with large FE problems with high-efficiency and high-accuracy. As a result, high-performance parallel FEM has become one of the essential way to solve practical engineering problems.

The challenge is how to create an parallel FEM system with higher cost performance. Unfortunately, most existing parallel FE methods are so complicating that they are not suitable for research. Therefore, MiniFE, which is simple but still covers all the important performance features of parallel FEM, is the best choice to study parallel finite element problems.

Based on MiniFE source codes, this work summarizes concrete steps, key computing pattern and parallel mechanism of parallel FEM. According to experimental results, we analyze the main performance bottleneck of the program. Based on that, we optimize the bottleneck of the MiniFE program - SpMV kernel, and provide an improving plan of hybrid MPI+OpenMP.

## 1 MiniFE source program analysis

MiniFE(Version 1.4) is a small test program for computer performance measurements, which comes from the Mantevo project of Sandia National Laboratories in National Energy Research Scientific Computing Center(NERSC). It mainly deals with given finite element problems.

MiniFE is self-contained independent code, which executes the whole FE phases —FE generation, FE assembly and FE analysis. The physical domain is a 3-D box simulated by hexahedral element. The box is discretized into structured mesh while it is regarded as unstructured mesh, then divide the domain through recursive coordinate bisection(RCB) to support parallel execution[2]. It employs linear FEM to generate sparse linear equation system from steady-state heat conduction equations on 3-D brick area, then solves this sparse linear equation system through conjugate gradient(CG) method without preprocessing. It contains 4 core parts as follows[5]:

● Element operator computing: generate "element equation system" for every element in the domain.

● Assemble "element equation system": assemble all the elements' "element equation system" to generate the final symmetric sparse linear equation system which needs to be calculated.

● Sparse matrix vector multiplication: it is used in the calculation of conjugate gradient method.

● Vector operation: including vector dot product, vector addition, etc. Used in the calculation of conjugate gradient method.

The problem domain of MiniFE is a unit space of hexahedral elements (0<=x<=1.0, 0<=y<=1.0, 0<=z<=1.0). And the known boundary conditions are: the value on the plane x=1.0 is 1.0; the value on the other boundary planes is 0.0. According to the finite element mesh size of every direction of input, i.e Nx, Ny and Nz, we can approximately work out the steady temperature value of each point in space. At the end of the program, we will compare the numerical solution which is calculated by FEM with analytical solution to verify the feasibility of FEM to solve this kind of problem. Fig 1.1 is the flow diagram of MiniFE program.

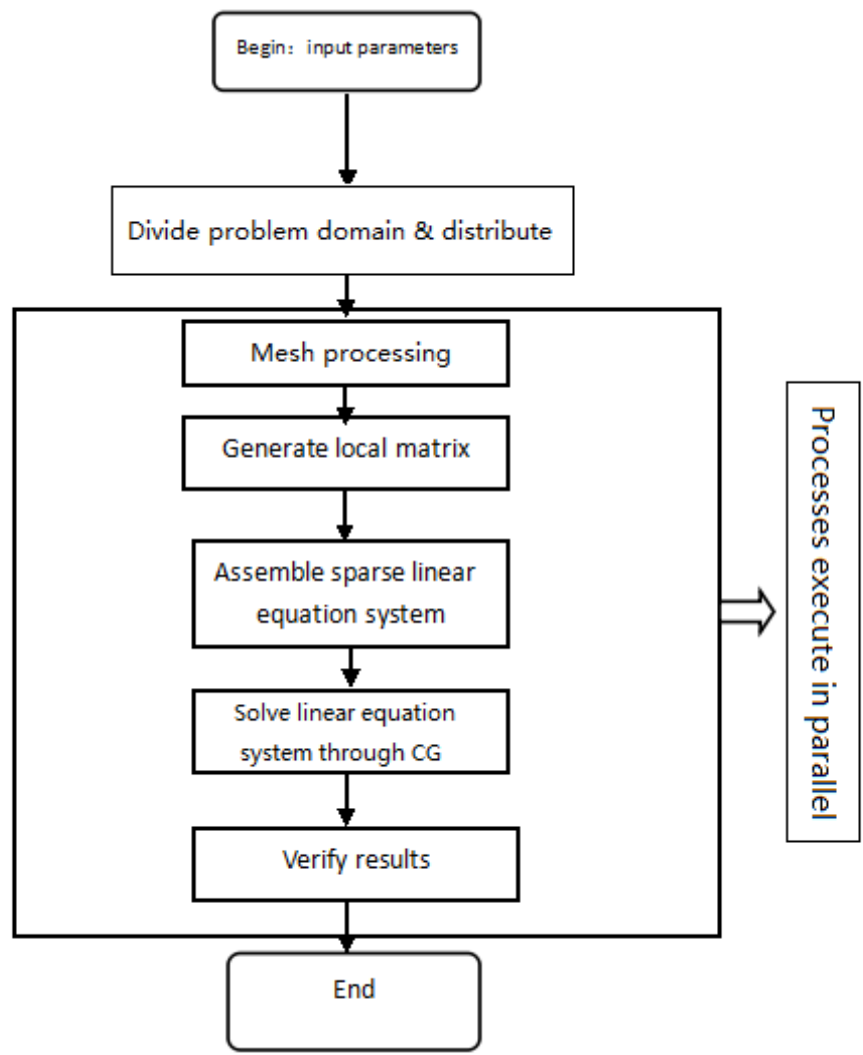

Fig 1.1 The flow diagram of MiniFE program

Fig 1.1 shows that MiniFE program contains several modules including problem domain partition & distribution, mesh processing, matrix generation, FE assembly, linear equation system solving with CG, results verification. All the modules are implemented in parallel by multiple processes except the problem domain partition module. At first, process 0 divide the problem domain into small regions according to the input parameters Nx, Ny, Nz. Then, distribute all the small regions to several processes. Next, every process implements the operations in following modules for their own small region. Inter-process communication is conducted by massage passing interface(MPI). The main function of each module is as follows:

● Problem domain partition & distribution module. Based on input parameters Nx, Ny, Nz and the specified process number, this module deals the the problem domain partition and distribution. Partition uses cursive coordinate bisection(RCB) to support the following distribution operation and parallel execution.

● Mesh processing module. After every process gets their own sub-domain, conduct corresponding mesh processing for each sub-domain, including store and mark the range of global area and local area, record the global row number of boundary condition nodes in current process' sub-domain, and map local nodes to global row number. These work prepare for the future calculation and matrix boundary value.

● Matrix generation module. This module conducts initialization operation for local sparse matrix A and determines the size of local sparse matrix. We use compressed row storage(CRS) to store sparse matrix.

● Finite element assembly module. This module deals with FE assembly and fills in "packed_coefs" array and "b" array of sparse matrix A. These operations are mainly conducted on each computing cell which is stored using data structure "ElemData" .

● Linear equations solving with CG module. This module uses conjugate gradient(CG) method to solve the sparse linear equation system A*x=b which is obtained by modules above. The result, vector x, is the numerical solution of space steady-state heat conduction equations. This module mainly contains 3 operations[6,7]: vector update(WAXPY), vector dot product(DOT) and sparse matrix vector multiplication(MATVEC).

● Result verification module. This module mainly calculates the difference between numerical solution and analytical solution.

## 2 MiniFE performance bottleneck analysis

Based on analysis of each module of MiniFE program described above, we carry out some experiments on a server platform to test MiniFE program. In the experiment, we set the input parameter Nx as 200/300 respectively, and processes number np as 1/4/8/16 respectively. The performance of program is measured by total execution time. Fig 2.1 and Fig 2.2 show the analysis conclusion of the experimental data.

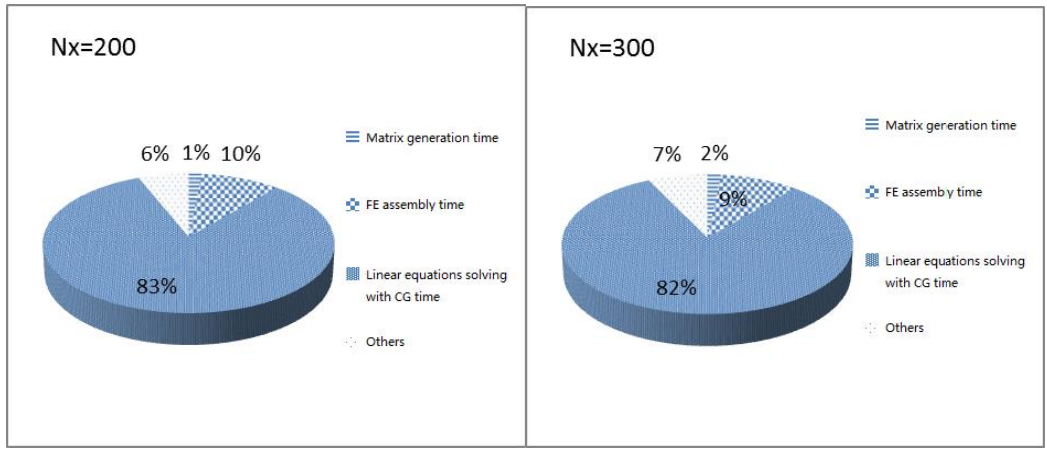

Fig 2.1 Nx=200/300, proportion of each module of the program

Fig 2.1 shows that:

● Linear equations solving with CG module consumes the largest part of MiniFE execution time. The proportion is around 82%-83%.

● The proportion of execution time of FE assembly module ranks second among all the modules, which is about 9%~10%.

● The proportion of execution time of matrix generation module is about 1%~2%.

● The rest parts consume 6%-7% of the execution time.

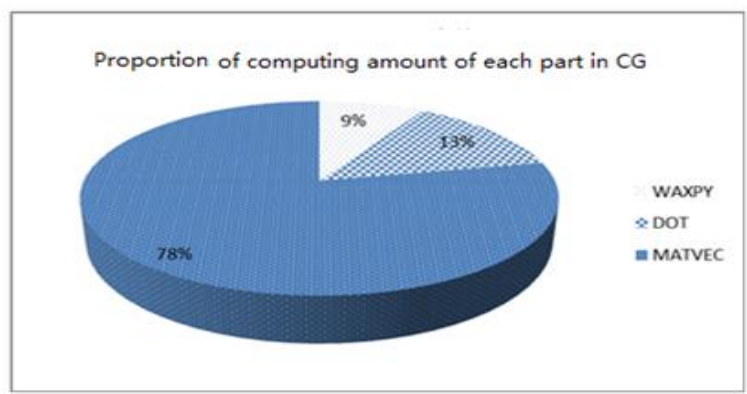

Fig 2.2 Proportion of computing amount of each part in CG method

Fig 2.2 shows that the main part of calculation in CG is MATVEC, i.e SpMV, and its proportion is about 78% while the proportion of vector update and vector dot product are respectively 9% and 13%. Therefore, the major performance bottleneck of this program is SpMV. So the efficient way to optimize MiniFE is to optimize SpMV.

## 3 MiniFE performance optimization

### 3.1 Optimization of SpMV

Sparse matrix vector multiplication(SpMV) is an important computing kernel which is widely used in scientific computing and practical application such as image processing, signal processing, iteration algorithm, etc. However, it is regarded as one of the "Seven Dwarfs" in the science and engineering computing filed[4]. Compared with dense matrix vector multiplication, performance of SpMV is very bad. Performance of SpMV in traditional storage mode is always lower than the peak value of machine's floating point operation by 10%[8]. MiniFE employs compressed row storage(CRS) to store sparse matrix.

In CRS, every non-zero value in sparse matrix A , every non-zero value's column number and the index of the first non-zero value in each row(if the row has at least one non-zero value) need to be stored. That is, 3 arrays need to stored. We assume that matrix A has m rows and n columns, and it contains nnz non-zero values. 3 arrays are defined as below:

- val [nnz]: record every non-zero value in A.

- col [nnz]: record every non-zero value's column number.

- ptr [m+1]: record the index of each row's first non-zero value in array val[nnz] and array col[nnz], and ptr[m]=nnz.

Fig 3.1 shows core codes of Matrix storage format of CRS in SpMV(y=A*x):

```
for(int i=0; i<m; ++i) {
        for(int j=ptr[i]; j<ptr[i+1]; ++j) {
          y [i] +=val[j]*x[col[j]];
        }
}
```

Fig 3.1 SpMV based on CRS storage format

There are 4 performance bottlenecks in CRS:

    1. When the sparse matrix has a very large size, the sequential access to array val or col may cause cache miss.

    2. Memory access to vector x lacks of spatial and temporal locality.

    3. Using array col to access array x needs indirect addressing.

    4. Every multiple-add operation needs to load 3 floating point number. As a result, memory access performance of CPU is the key point.

Focused on indirect memory access problem in CRS, we chose more efficient data compression method to improve program performance. Since storage format of CRS is more flexible and efficient during matrix generation and FE assembly module, we still use CRS for matrix generation and storage before solving linear equation system with CG. At the begin of solving linear equations with CG module, we change the storage format of sparse matrix, CRS, into another storage format which is more suitable for iteration. Then, we start to solve equations with CG.

In MiniFE program, the sparse matrix generated by FEM is a symmetric banded matrix. Non-zero values of each row in this matrix show a trend that several of them are always adjacent, thus their column numbers have spatial locality. Therefore, we can store these adjacent non-zero values in the same row together as a non-zero segment. This kind of storage mode not only reduces the size of memory space, but also decreases number of memory access. This strategy improves the computing to memory access ratio of SpMV, thus improving performance of SpMV.

We choose specific block compressed row storage(BCRS)[8] instead of CRS to store matrix. In BCRS, size of sub-block is 1*n, and the value of n is variable. n in each sub-block equals to the length of the corresponding non-zero segment. There are 4 arrays need to be stored in BCRS. We assume that matrix A has m rows and n column, and has nnz non-zero value sand nns non-zero segments. 4 arrays are defined as below:

    ● val[nnz]: record every non-zero value.

    ● jas[nns]: record the column number of the first non-zero value in every non-zero

segment.

- offsets[nns+1]: record the index of every non-zero segment's first non-zero value in array val[nnz], offsets[nns]=nnz.

- ptr[m+1]: record the index of every row's first non-zero segment in array jas[nns] and offsets[nns+1], ptr[m]=nns.

Fig 3.2 shows core codes of Matrix storage format of BCRS in SpMV(y=A*x).

```
int col;
for(int i=0;i<m;i++){
    double sum=0;
      for(int j=ptr[i];j<ptr[i+1];j++){
         col=jas[j];
         for(int k=offsets[j];k<offsets[j+1];k++){
           sum+=val[k]*x[col++];
         }
      }
       y[i]=sum;
}
```

Fig 3.2 SpMV based on BCRS storage format

As shown in Fig 3.2, innermost loop of SpMV based on BCRS is too short that it will cause a lot of loop overhead. Moreover, every loop needs branch decision which is likely to cause pipeline stalls, thus limiting the improvement of performance. After analyzing the feature of the matrix, we find that : after inter-process communication processing on the matrix generated by FEM, the length of non-zero segment (i.e the number of non-zero value in the segment) must be 1, 2, 3 or 4. And non-zero segments of length 3 account for most of them. Therefore, we replace the innermost "for" loop with "case" statement to implement loop unrolling. We deal with non-zero segment of length 3 first to reduce the number of loop decision. Fig 3.3 shows the improved algorithm.

```
int col,n,k,length;
for(int i=0;i<m;i++){
    double sum=0;
     for(int j=ptr[i];j<ptr[i+1];j++){
        col=jas[j];
        n=offsets[j];
        k=offsets[j+1];
        length=k-n;
        switch(length){
        case 3: (EXECUTE 3 TIMES)   sum+=val[n++]*x[col++];
                break;
        case 1: (EXECUTE 1 TIME)   sum+=val[n++]*x[col++];
                break;
        case 2: (EXECUTE 2 TIMES)   sum+=val[n++]*x[col++];
                break;
        case 4: (EXECUTE 4 TIMES)   sum+=val[n++]*x[col++];
                break;
        }
    }
      y[i]=sum;
}
```

Fig 3.3 Improved SpMV based on BCRS

**Complexity analysis of SpMV based on CRS and BCRS:**

● Computational complexity: computational complexity of CRS is $O(2*nnz)$; every non-zero segment needs one more calculation in BCRS, thus its computational complexity is $O(2*nnz+nns)$.

● Memory access complexity: it will access array val nnz times, array col nnz times, array x nnz times, array y m times, array ptr m+1 times during one SpMV of CRS execution cycle, thus the memory access complexity is $O(3nnz+2m+1)$; it will access array val nnz times, array x nnz times, array jas nns times, array offsets nns+1 times, array y m times, array ptr m+1 times during one SpMV of BCRS execution cycle, thus the memory access complexity is $O(2nnz+2nns+2m+2)$.

We can conclude that calculations in BCRS is nns times more than in CRS while memory access in BCRS is nnz-2*nns times less than in CRS during one SpMV execution cycle. Because the length of non-zero segment in sparse matrix generated by FEM is mostly 3, nnz is approximately equal to 3 times of nns. So nnz-2*nns is approximately equal to nns. The limited bandwidth of SpMV leads to excess of CPU computing power. BCRS runs more calculation and less memory access, hence increasing the computing to memory access ratio of SpMV. Therefore, we can improve the performance of SpMV and shorten run time of the program.

**Analysis of memory capacity of SpMV based on CRS and BCRS:**

● CRS needs to store 3 arrays, and the total memory capacity is nnz*sizeof(val[0])+ nnz*sizeof(col[0])(m+1)* sizeof(ptr[0]).

● BCRS needs to store 4 arrays, and the total memory capacity is nnz*sizeof(val[0])+ nns*sizeof(jas[0])+(nns+1) * sizeof(offsets[0])+(m+1)* sizeof(ptr[0]).

Since the storage type of array col is the same as array jas, and nnz is approximately equal to 3*nns, the memory space of BCRS storage matrix is less than CRS storage matrix by nns*sizeof(col[0]) when SpMV is running. This part of memory can be provided to other programs on the experiment platform, as a consequence, it decreases its occupation of resources of the experiment platform. Moreover, compared to array col, array jas and array offsets need to store less data, so it can reduce cache misses during memory access in a certain degree and improve the performance of the program.

In the experiment, we set parameter nx as 200/300, process number np as 1/4/8/16 respectively to test the performance of the original program and the improved program. Fig 3.4 and Fig3.5 show the experimental results after we carefully analyze the experimental data.

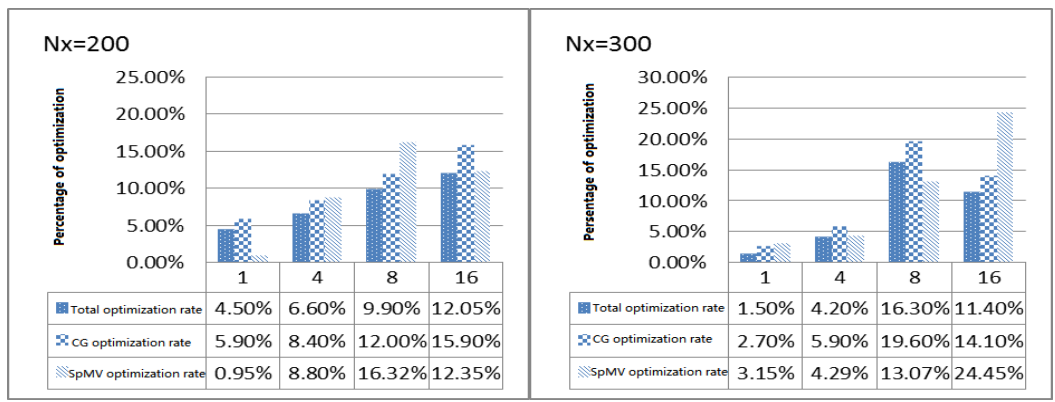

Fig 3.4 Percentage of each part's optimization in BCRS

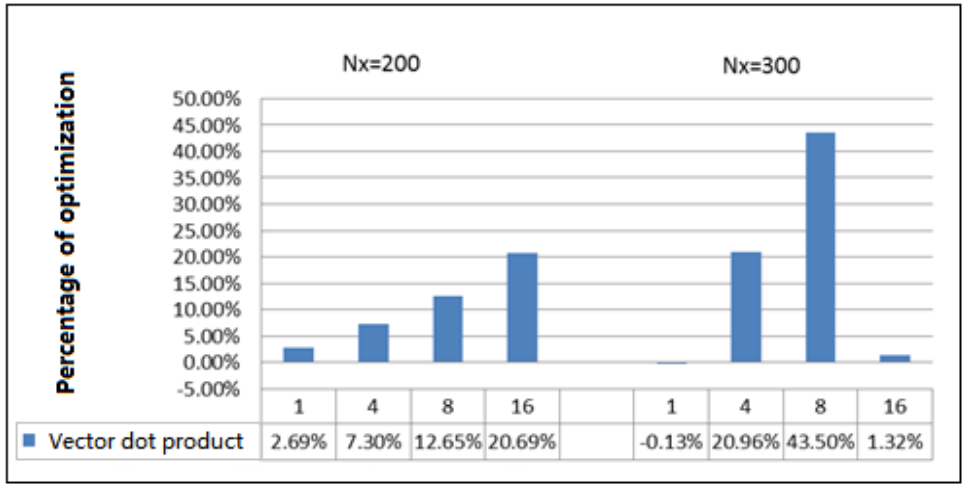

Fig 3.5 Percentage of optimization of vector dot product in CG part in BCRS

As shown in Fig 3.4, after optimization for storage, performance of some parts of SpMV is actually improved, and the optimization rate is increasing with the increase of process number. Besides, optimization rate of CG is higher than the rate of SpMV in the same group of experimental data. So we can assert that there are other optimization in CG except for optimization for SpMV. Fig 3.5 shows that performance of the vector dot product part in CG is also improved to some extent. The reason is as follows:

When program is running in parallel, the CG part executes vector dot product operation right after executing SpMV. Vector dot product operation needs a global communication to obtain global synchronization. When applying BCRS, every process does not need to access to array col during executing SpMV. Non-zero segment supports to access to the column in the segment one by one, thus using new calculation to replace old memory access. Compared with performance of memory access to arrays, computing performance of each process is more balanced and execution time is closer. In the original memory access, cache miss of some process may cause other processes' waiting, and global communication can not perform in time, which causes running time of dot product operation increases. After storage format is improved, execution time of each process is very close, which means that we can carry out global communications next to local dot product operation. Therefore, execution time of vector dot product is shorter than the original execution time. In a word, BCRS storage structure makes load of each process well balanced. Moreover, array jas and array offsets in this structure are relatively short. Compared to access to array col in the original program, the improved way can reduce cache miss and enhance synchronization.

After analyzing and synthesizing experimental results in different data scale and different process amount, we can obtain the optimization rate of performance of the improved program as follows:

● Performance of SpMV is improved by 24.45% in the best case, and improved by 0.95% in the worst case. The average optimization rate of SpMV is 10.42%.

● Performance of vector dot product is improved by 43.50% in the best case, and lowered by 0.13% in the worst case. The average optimization rate of vector dot product is 13.63%.

● Performance of CG is improved by 19.60% in the best case, and improved by 2.70% in the worst case. The average optimization rate of CG is 10.56%.

● Performance of the whole program is improved by 16.30% in the best case, and improved by 1.50% in the worst case. The average optimization rate of the program is 8.31%.

### 3.2 Optimization for MPI+OpenMP parallel programming

MiniFE program has both MPI parallel programming with global distributed memory and OpenMP parallel programming with local shared memory. This mix of MPI and OpenMP programming can make full use of system resources and improve parallel performance of the program. However, the MPI programming part still can be further improved; the OpenMP part is only used in parallel optimization of CG module. Therefore, we can complete MPI+OpenMP parallel programming in the whole program and further improve performance of the program.

● Parallel optimization of MPI

In some global communication of MPI parallel programming, only process 0 will use the final data while other processes do not need to store these data. Thus some operation which needs global communication only needs to transfer data to process 0 instead of all the processes.

Therefore, we can change some appropriate MPI_Allgather( ) into MPI_Gather( ) and change MPI_Allreduce( ) into MPI_Reduce( ). So we can reduce communication overhead to a certain degree, speed up each process in the program and improve the program's performance.

⚫ Parallel optimization of OpenMP

```
#pragma omp parallel for
   for(int sz=-1; sz<=1; ++sz) {
           for(int sy=-1; sy<=1; ++sy) {
             for(int sx=-1; sx<=1; ++sx) {
               GlobalOrdinal col_id =
get_id<GlobalOrdinal>(global_nodes_x,global_nodes_y,global_nodes
_z, ix+sx, iy+sy, iz+sz);
               if (col_id >= 0 && col_id < global_nrows) {
                   #pragma omp actomic
                 ++nnz;
               }
             }
           }
   }
```

Fig 3.6 Optimization of OpenMP in matrix generation module

Fig 3.6 shows the optimization of OpenMP in matrix generation module. In this module, we can apply OpenMP to generation part of non-zero value's column number of the matrix to realize parallel optimization. Similar optimization can also be applied to domain partition module.

```
#pragma omp parallel for firstprivate(elem_data)      //elem_data is defined as
    ▪  for(size_t i=0; i<elemIDs.size(); ++i) {         //a local private variable
    ▪  get_elem_nodes_and_coords(mesh, elemIDs[i], elem_data);
    ▪  compute_element_matrix_and_vector(elem_data);
       #pragma omp critical{
    ▪  sum_into_global_linear_system(elem_data, A, b);
    ▪  }
   }
```

Fig 3.7 Optimization of OpenMP in FE assembly module

Fig 3.7 shows optimization of OpenMP in FE assembly module. In this module, we can apply OpenMP to calculation during computing cell assembled into the total stiffness matrix. Because every process needs to use elem_data independently, we define it as a local private variable. When adding matrix and vector part of current computing cell into global system A and b respectively, each process will compete for matrix A and vector B, so a critical area #pragma omp critical{} is needed.

Optimization for MPI+OpenMP hybrid programming is described as above. Experimental data shows that these modification can increase performance of the program to a certain extent when data size is large enough, but the results of improvement are not obvious. This is related to the specific experimental platform. Parallel overhead in a server platform is relatively large. However, if the program is running in some platform such as GPU which is more efficient for fine-grit multithreading parallel, performance of the program may increase a lot.

## 4 Conclusion

This work uses MiniFE source codes as research basis, introduces background knowledge about the program, analyzes implementation method of each module in the program, and focuses on implementation steps, kernel calculation mode and parallel computing feature of FE parallel computing. Through our experiment, we obtain the the proportion of calculation amount of each module and we find that the main performance bottleneck of the program is sparse matrix vector multiplication(SpMV). Based on that, we choose a proper platform to conduct the experiments. In order to reduce access times and improve computing to memory access ratio of SpMV, considering feature of the sparse matrix generated by MiniFE program, we use specific block compressed row storage(BCRS) to store parse matrix. We compare BCRS with the original storage mode CRS, and decide that CRS or BCRS should be applied to different appropriate stages of the program. We also implement programming of SpMV using BCRS mode. In addition, we analyze relevant performance indicators of the original program and the improved program in different experiment parameters. The improved program can provide on average 8.31% higher performance of the total program.

Moreover, an improving plan of hybrid MPI+OpenMP programming is provided which can further promote improvement of the performance of MiniFE.

# References

[1] 陈全，诸昌铃，张本才.并行有限元计算的一种新途径[J].路基工程， 2009（3）.

[2] Carey G F.Parallelism in finite element modeling[J].Commun Appl,Numer,Meth,1986,2:281-287.

[3] http://www.nersc.gov/systems/nersc-8-procurement/trinity-nersc-8-rfp/nersc-8-trinity-benc%20hmarks/minife

[4] 王迎瑞，任江勇，田荣.基于 GPU 的高性能稀疏矩阵向量乘及 CG 求解器优化[J].计算机科学，2013，40（3）.

[5] 袁娥.稀疏矩阵向量乘的存储访问复杂度及自动性能优化技术研究.中国科学院软件研究所硕士学位论文,2008.

[6] 裴琴娟.解线性方程组的共轭梯度法[J].新乡学院报：自然科学版，2011，（28）：309.

[7] 王迎瑞，任江勇，田荣.基于 GPU 的高性能稀疏矩阵向量乘及 CG 求解器优化[J].计算机科学，2013，40（3）.

[8] 袁娥.稀疏矩阵向量乘的存储访问复杂度及自动性能优化技术研究.中国科学院软件研究所硕士学位论文,2008.